\documentclass[aps,showpacs,superscriptadress,preprint]{revtex4}
\usepackage{graphicx}
\usepackage{amsmath}
\usepackage{amssymb}
\usepackage{mathrsfs}
\usepackage{stmaryrd}
\usepackage{amsthm}
\usepackage{hyperref}
\hypersetup{hypertex=true,
	colorlinks=true,
	linkcolor=blue,
	anchorcolor=blue,
	citecolor=blue}
\usepackage{caption}
\usepackage{tikz,xcolor,hyperref}

\definecolor{lime}{HTML}{A6CE39}
\DeclareRobustCommand{\orcidicon}{
\begin{tikzpicture}
\draw[lime, fill=lime] (0,0)
circle[radius=0.16]
node[white]{{\fontfamily{qag}\selectfont \tiny \.{I}D}}; 
\end{tikzpicture}
\hspace{-3mm}
}
\foreach \x in {A, ..., Z}{%
\expandafter\xdef\csname orcid\x\endcsname{\noexpand\href{https://orcid.org/\csname orcidauthor\x\endcsname}{\noexpand\orcidicon}}
}

\begin{document}
\title{Shadows and quasinormal modes of a charged non-commutative black hole by different methods}

\author{Zening Yan$^{1}$\footnote{Corresponding author: z.n.yan.bhtpr@gmail.com}\hspace{-2mm}\orcidA{}, Xiaoji Zhang$^{2}$, Maoyuan Wan$^{2}$ and Chen Wu$^{3}$} 

\affiliation{
\small 1. Department of Physics and Center for Field Theory and Particle Physics, Fudan University, Shanghai 200433, China\\
\small 2. College of Science, University of Shanghai for Science and Technology, Shanghai 200093, China\\
\small 3. Shanghai Advanced Research Institute, Chinese Academy of Sciences, Shanghai 201210, China\\}

\date{July 9, 2021}

\begin{abstract}
In this paper, we calculated the quasinormal modes (QNMs) of a charged non-commutative black hole in scalar, electromagnetic and gravitational fields by three methods.
We gave the influence of non-commutative parameter $\theta$ and charge $Q$ on QNMs in different fields.
Thereafter, we calculated the shadow radius of the black hole and provided the valid range of $\theta$ and $Q$ using the constraints on the shadow radius of $\text{M87}^{\ast}$ and $\text{Sgr\;A}^{\ast}$ from the Event Horizon Telescope (EHT).
In addition, we estimated the ``relative deviation'' of the shadow radius ($\delta_{R_{s}}$) between non-commutative spacetime and commutative spacetime.
We found that the maximum values of $\delta_{R_{s}}$ decreases with the increase of charge $Q$.
In other words, the non-commutativity of spacetime becomes harder to distinguish as the charge of the black hole increases.
\end{abstract}


\maketitle

\section{Introduction and motivation}
In 1947, Snyder published the first paper on the non-commutativity of spacetime  \cite{Snyder:1946qz}. 
Since then, non-commutative geometric black holes have been studied more deeply. 
At present, there are two different methods to study non-commutative quantum field theory: the Weyl-Wigner-Moyal $\star$-product \cite{Groenewold:1946kp, Moyal:1949sk} and the other on coordinate coherent state formalism \cite{Smailagic:2003yb, Smailagic:2003rp, Smailagic:2004yy}.
The so-called `` $\star$-product '' method is that the ordinary product between functions is replaced by the $\star$-product
\begin{equation} 
f(x) \star g(x)\equiv\left.e^{\frac{i}{2} \theta^{\mu \nu} \frac{\partial}{\partial x^{\mu}} \frac{\partial}{\partial \nu^{\nu}}} f(x) g(y)\right| _{y \rightarrow x},
\end{equation}
where $f(x)$ and $g(x)$ are two arbitrary infinitely differentiable functions.
It can be seen that the non-commutativity of spacetime is contained in $\star$-product.
The coordinate coherent state method means that the Dirac $\delta$ distribution in ordinary spacetime is replaced by Gaussian distribution.
In the framework of this method, Nicolini believes that the non-commutative effect only acts on the material source term and that there is no need to change the Einstein tensor part of the field equation \cite{Nicolini:2005vd}.
Specifically, the mass density of the point-like function on the right hand side of the Einstein field equation is replaced with the Gaussian smeared matter distribution, while the left hand side of the equation is unchanged.

Nicolini first proposed the four-dimensional non-commutative - geometry - inspired Schwarzschild black hole \cite{Nicolini:2005vd}.
After that, this black hole solution was extended to the case of charged \cite{Ansoldi:2006vg}.
In 2010, Modesto and Nicolini extended it to the general case of charged rotating non-commutative black holes \cite{Modesto:2010rv}. 
On this basis, many researchers have explored the practical application of this black hole, such as Douglas M. Gingrich, who explores the possibility of the production and decay of non-commutative-geometry-inspired black holes on the Large Hadron Collider (LHC) at the phenomenological level \cite{Gingrich:2010ed}.
Chikun Ding and Jiliang Jing studied the influence of non-commutative parameter on strong field gravitational lensing in the non-commutative Reissner-Nordstr$\ddot{\text{o}}$m black-hole \cite{Ding:2011az}.
In addition, some authors use the observed data of gravitational waves and binary pulsars to discuss the constraint of quantum fuzziness scale of non-commutative spacetime and non-commutative gravity \cite{Kobakhidze:2016cqh, Jenks:2020gbt}.

On the other hand, the QNMs has always been an attractive topic in the research field of black holes \cite{Konoplya:2011qq, Berti:2009kk}. 
The QNMs is the characteristic oscillations of the matter field under the background of spacetime.
The numerical result of the perturbation frequency is a complex number, which has been found by the LIGO Scientific Collaboration when detecting the gravitational wave generated by the merger of two black holes \cite{LIGOScientific:2016aoc}. 
The reason why people are interested in the QNMs stage is that it can be seen as a ``characteristic sound'' of a black hole. 
This characteristic oscillation does not come from the matter inside the black hole, but mostly depends on the spacetime metric outside the event horizon of the black hole. 
This shows that spacetime itself is also a direct participant in the oscillation.

The shadow of a black hole is a two-dimensional dark area formed on the celestial sphere under the influence of a strong gravitational effect.
In 2000, Falcke and his collaborators first proposed that the shadow of a black hole could be observed \cite{Falcke:1999pj}. 
In 2019, the EHT first showed the image of the supermassive black hole $\text{M87}^{\ast}$ \cite{EventHorizonTelescope:2019dse, EventHorizonTelescope:2019ggy, EventHorizonTelescope:2019pgp}, from which shadows can be clearly observed.
In 2022, the EHT released an image of a supermassive black hole $\text{Sgr\;A}^{\ast}$ at the center of the Milky way galaxy \cite{EventHorizonTelescope:2022xnr, EventHorizonTelescope:2022vjs, EventHorizonTelescope:2022wok, EventHorizonTelescope:2022exc, EventHorizonTelescope:2022urf, EventHorizonTelescope:2022xqj}.
In general, the shadow cast by a non-rotating spherically symmetric black hole is a circle, but for a rotating black hole, its shadow is similar to the shape of the letter ``$\text{D}$'' \cite{Perlick:2021aok, Chael:2021rjo, Gralla:2019xty, Falcke:1999pj}.

The motivation of this paper is listed as follows:
1) In this paper, we will test whether the corresponding relationship between eikonal perturbation and shadow radius is valid in the case of non-vacuum Einstein's equation solution;
2) In some papers, the WKB method fails to obtain correct numerical results when calculating the QNMs of the non-commutative black hole. Therefore, we will use other methods to calculate and give the correct numerical results of QNMs in scalar, electromagnetic and gravitational fields;
3) We will use the constraint range of shadow radius from EHT to limit the range of $\theta$ and $Q$.

We organize the paper as follows:
In Section 2, 
we introduce the basic equations used to calculate the perturbation. And the valid range of $\theta$ corresponding to different $Q$ values is calculated in the case of ensuring the existence of the event horizon of the black hole.
In Section 3,
the QNMs of charged non-commutative black holes in different fields are calculated by three methods.
In Section 4, 
We test the relationship between shadow radius and eikonal perturbation in charged non-commutative black hole spacetime, and calculate the shadow radius of black hole corresponding to different parameters.
In Section 5, 
a brief summary of the whole paper is given.
We use natural units $\left(G=c=\hbar=1\right)$ throughout the paper.

\section{The basic equations}
\subsection{The metric of charged non-commutative black hole spacetime}
When we define the spacetime coordinate $X^{\mu}$, non-commutative spacetime can be expressed as
\begin{equation} 
\left[X^{\mu}, X^{\nu}\right]=i \theta^{\mu \nu}=i \theta \epsilon^{\mu \nu}.
\end{equation}
The non-commutative parameter $\theta$ is a positive number and its dimension is $length^{2}$, $\epsilon^{\mu \nu}$ is the anti-symmetric tensor \cite{Nicolini:2008aj}.
We use the non-commutative spacetime metric constructed by coordinate coherent state, that is, the ``point particle'' mass in spacetime is no longer described by Dirac $\delta$ function, but by Gaussian function instead of matter distribution.
Because there is no point object, no physical distance can be less than the minimum position uncertainty of the order of $\sqrt{\theta}$ \cite{Spallucci:2008ez}.

The contours of the Gaussian distribution corresponding to matter and charge density are
\begin{equation}
\rho_{\text{matt}}(r)=\frac{\mathcal{M}}{(4 \pi \theta)^{3 / 2}}  \text{e}^{-\frac{r^{2}}{4 \theta}}, \quad  \rho_{\text{el}}(r)=\frac{\mathcal{Q}}{(4 \pi \theta)^{3 / 2}}  \text{e}^{-\frac{r^{2}}{4 \theta}},
\end{equation}
where $\mathcal{M}$ is the ``bare mass'' and $\mathcal{Q}$ is the total electric charge.
Here we need to consider a quasi-classical system, analogous to the traditional Einstein-Maxwell system, which describes both electromagnetic and gravitational fields \cite{Nicolini:2008aj}
\begin{equation}
\mathcal{R}^{\mu}{ }_{\nu} - \frac{1}{2} \delta^{\mu}{ }_{\nu} \mathcal{R} = 8 \pi\left( \left.T^{\mu}_{\nu}\right|_{\text{matt}} + \left.T^{\mu}_{\nu}\right|_{\text{el}} \right),
\end{equation}
\begin{equation}
\frac{1}{\sqrt{-g}} \partial_{\nu} \Big( \sqrt{-g} \delta^{0[\mu|} \delta^{r|\nu]} E(r) \Big)=J^{\mu}, \quad  J^{\mu}(x)=4 \pi \rho_{\text{el}}(r) \delta_{0}^{\mu}.
\end{equation}
The tensors $\left.T^{\mu}_{\nu}\right|_{\text{matt}}$ and $\left.T^{\mu}_{\nu}\right|_{\text{el}}$ are the energy momentum tensors, describing the matter and the electromagnetic content.
$E(r)$ is the the electric field and $J^{\mu}$ is the corresponding current density.

The line element for a static spacetime with spherical symmetry can be written as follows:
\begin{equation}\label{metricofsss}
ds^2= g_{\mu \nu} d x^{\mu} d x^{\nu}=-A(r) d t^{2}+B(r) d r^{2}+D(r)\left(d \vartheta^{2}+\sin ^{2} \vartheta d \varphi^{2}\right),
\end{equation}
where the spherical coordinate $(t,r,\vartheta,\varphi)$ is used.
This metric satisfies $A(r)=B(r)^{-1}=f(r)$, $D(r)=r^2$, for the charged non-commutative spacetime, and the lapse function \cite{Ansoldi:2006vg, Nicolini:2008aj, Ding:2011az} is given by
\begin{equation}\label{Lapsefunction}
f(r)=1-\frac{4M}{r\sqrt{\pi}}\gamma\left(\frac{3}{2},\frac{r^2}{4\theta}\right)
+\frac{Q^2}{r^2\pi}\left[\gamma^2\left(\frac{1}{2},\frac{r^2}{4\theta}\right) -\frac{r}{\sqrt{2\theta}}\gamma\left(\frac{1}{2},\frac{r^2}{2\theta}\right) +r\sqrt{\frac{2}{\theta}}\gamma\left(\frac{3}{2},\frac{r^2}{4\theta}\right) \right] \text{,}
\end{equation}
where the lower incomplete gamma function is written as
\begin{equation}
\gamma(s, x)=\int_{0}^{x} t^{s-1} \mathrm{e}^{-t} \mathrm{~d} t \text{,}
\end{equation}
and notice that
\begin{equation}
M=\frac{2 \mathcal{M}}{\sqrt{\pi}} \Gamma\left(\frac{3}{2}\right)+\sqrt{\frac{2}{\theta}} \mathcal{Q}^2  \Gamma\left(\frac{1}{2}\right), \quad  Q=\mathcal{Q}.
\end{equation}
When $r\rightarrow0$ or $r\rightarrow\infty$, the limit of the lapse function is
\begin{equation}
\lim _{r \rightarrow 0 \  \text{or} \  r \rightarrow \infty} f(r) \equiv1 \text{,}
\end{equation}

\begin{figure}[htbp]
\centering 
\includegraphics[height=4.62cm,width=15cm]{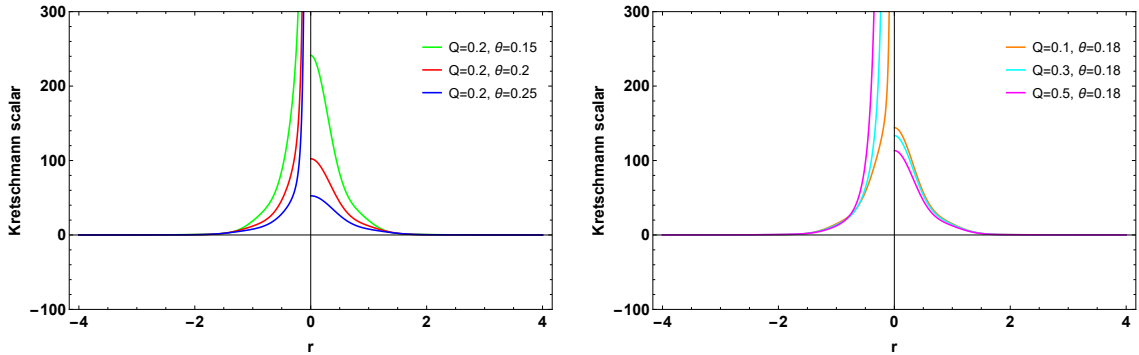}
\caption{The function of the Kretschmann scalar with $r$ as the independent variable. The parameter $M=1$ is selected.}\label{KS}
\end{figure}

In order to explore the singularity of the charged non-commutative black hole spacetime more rigorously, we calculated the Kretschmann scalar
\begin{equation}
\mathcal{K} = \mathcal{R}_{\mu \nu \rho \sigma} \mathcal{R}^{\mu \nu \rho \sigma}.
\end{equation}
Then we show the variation of the Kretschmann scalar with parameters $\theta$ and $Q$ in Fig. \ref{KS}.
We can see from the figure that the Kretschmann scalar does not diverge when $r\geq0$. 
However, the Kretschmann scalar diverges when the negative value $r$ approaches zero.
When $M=1$ and $r$ is an infinitesimal positive number $(r \rightarrow 0)$, the limit of the Kreichman scalar is
\begin{equation}
\lim _{r \rightarrow 0} \mathcal{K}(r) = \frac{Q^4 - 4 \sqrt{2 \pi \theta}  Q^2  + 8 \pi \theta}{3 \pi^2 \theta^4}.
\end{equation}
As a result, we discover that there are no singularities in the $[0, \infty)$ range of the charged non-commutative black hole spacetime.
Based on this, this study only discusses the situation within the range of non-singularity.

In addition, Eq. (\ref{Lapsefunction}) can be reduced to several well-known black hole solutions when the non-commutative parameter $\theta$ and the parameter $Q$ take the following limit cases:
$$
\left\{
\begin{aligned}
         \lim _{\theta \rightarrow 0} f(r) &= 1-\frac{2M}{r}+\frac{Q^2}{r^2} \;  & \text{Reissner-Nordstr$\ddot{\text{o}}$m (Singularity)} \\
	 f(r)|_{Q=0} &= 1-\frac{4M}{r\sqrt{\pi}}\gamma\big(\frac{3}{2},\frac{r^2}{4\theta}\big) \;  & \text{Non-commutative-geometry-inspired (No singularity)}          \\
		\lim _{\theta \rightarrow 0} f(r)|_{Q=0}& = 1-\frac{2M}{r} \; & \text{Schwarzschild (Singularity)}
\end{aligned}
\right. \text{.}
$$
It is well known that the charged non-commutative metric is a non-vacuum static solution of the Einstein-Maxwell field equations.

\begin{figure}[htbp]
\centering 
\includegraphics[height=5cm,width=15cm]{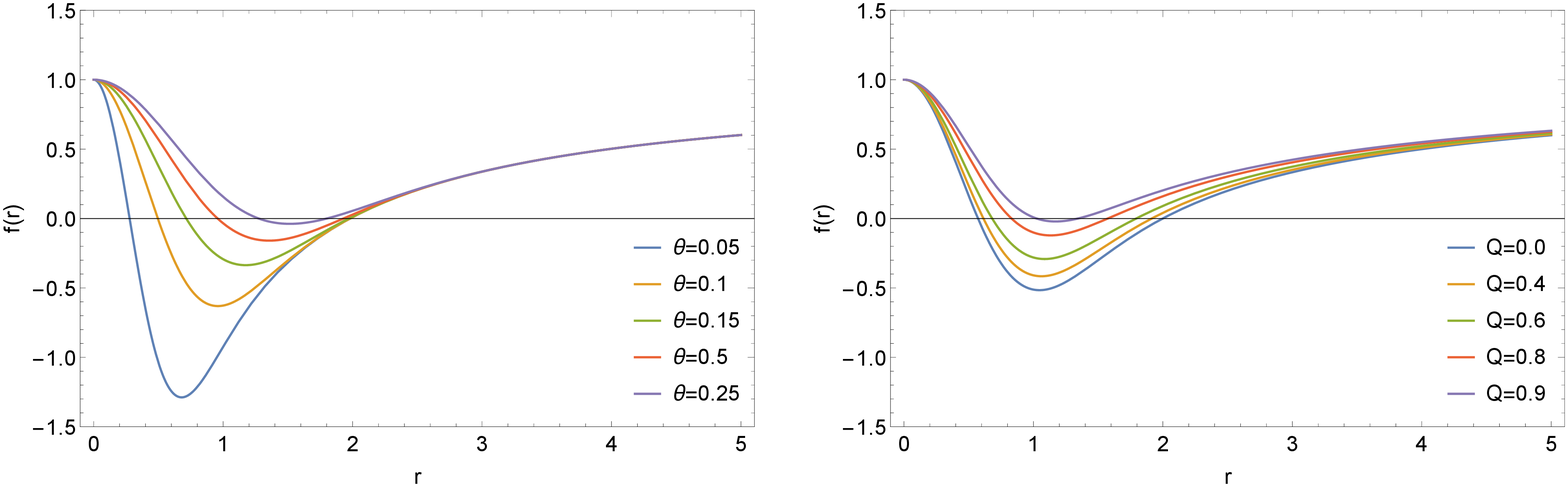}
\caption{The lapse functions of charged non-commutative black holes corresponds to different parameters. The parameters $Q=0.2$ (left panel), $\theta=0.12$ (right panel) and $M=1$ are selected.}\label{LF}
\end{figure}

Fig. \ref{LF} shows the change of the lapse functions of the charged non-commutative black hole with parameters $\theta$ and $Q$.
Obviously, it can be seen that there is a critical value $\theta_{max}$ for the non-commutative parameter $\theta$ to ensure that the black hole has at least one event horizon on the left panel of Fig. \ref{LF}.

\begin{figure}[htbp]
\centering
\includegraphics[height=10cm,width=12.5cm]{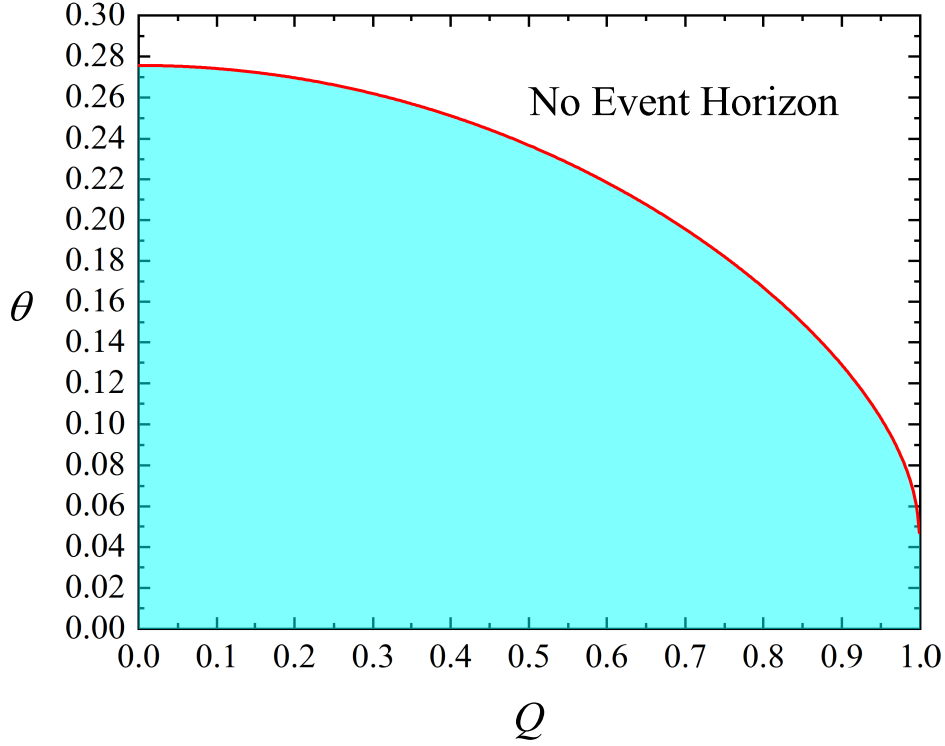}
\caption{The valid range of $\theta$ for various $Q$ values (cyan area) and the critical value $\theta_{max}$ (red line). The parameter $M=1$ is selected.}
\label{fig11}
\end{figure}

As shown in Fig. \ref{fig11}, when a black hole has only one event horizon, the corresponding critical value $\theta_{max}$ is calculated, implying that the effective range of $\theta$ is $(0, \theta_{max})$.
It can also be seen that the critical value $\theta_{max}$ decreases as $Q$ increases.

\subsection{Effective potential in the scalar, electromagnetic, and gravitational fields}
We consider the Klein-Gordon equation to solve the equation of massless scalar:
\begin{equation}\label{KG}
\qedsymbol \Psi=\frac{1}{\sqrt{-g}}\partial_\mu\big[\sqrt{-g}g^{\mu\nu}\partial_\nu\Psi \big]=0 \text{,}
\end{equation}
and assume that the scalar field has the following form:
\begin{equation}\label{Psi}
\Psi(t,r,\vartheta,\varphi)=\frac{1}{r}\sum_{l,m}e^{-i\omega t}\Phi(r)Y_{lm}(\vartheta,\varphi) \text{.}
\end{equation}	
Substituting Eq. (\ref{Psi}) into Eq. (\ref{KG}), we obtain the Schr$\ddot{\text{o}}$dinger-like wave equation:
\begin{equation}\label{wave1}
\frac{\partial^2\Phi(r)}{\partial r_*^2}+\big[\omega^2-V(r)_{scalar}\big]\Phi(r)=0 \text{,}
\end{equation}
where tortoise coordinate is $r_\ast= \int dr/f(r)$.
Therefore, the effective potential of scalar field is
\begin{equation}\label{Vs}
V(r)_{scalar}=f(r)\left[\frac{l(l+1)}{r^2} +\frac{1}{r}f^{\prime}(r) \right] \text{,}
\end{equation}
where $l$ represents the multipole number.

We consider the evolution of a Maxwell field \cite{Cardoso:2001bb}.
The covariant equation is given by
\begin{equation}\label{Maxwell}
\frac{1}{\sqrt{-g}}\partial_\mu\left[ \left(A_{\sigma,\delta}-A_{\delta,\sigma}\right) g^{\delta\nu}g^{\sigma\mu} \sqrt{-g} \right]=0 \text{.}
\end{equation}
The vector potential $A_\mu$ can be expanded into a four-dimensional vector spherical harmonics
\begin{equation}\label{harmonics}
A_\mu(t,r,\vartheta,\varphi)=\sum_{l,m}\begin{pmatrix} \begin{bmatrix} 0 \\ 0 \\ \frac{a^{lm}(t,r)}{\sin\vartheta}\partial_\varphi Y_{lm}(\vartheta,\varphi) \\ -a^{lm}(t,r)\sin\vartheta\partial_\vartheta Y_{lm}(\vartheta,\varphi) \end{bmatrix}+ \begin{bmatrix} f^{lm}(t,r)Y_{lm}(\vartheta,\varphi) \\ h^{lm}(t,r)Y_{lm}(\vartheta,\varphi) \\ k^{lm}(t,r)\partial_\vartheta Y_{lm}(\vartheta,\varphi)\\ k^{lm}(t,r)\partial_\varphi Y_{lm}(\vartheta,\varphi) \end{bmatrix} \end{pmatrix} \text{,}
\end{equation}
the first term has parity $(-1)^{l+1}$ and the
second $ {(-1)}^l $.
Substituting Eq. (\ref{harmonics}) into Eq. (\ref{Maxwell}), the wave equation can be obtained as follows
\begin{equation}\label{wave2}
\frac{\partial^2\psi(r)}{\partial r^2_*}+[\omega^2-V(r)_{ele}]\psi(r)=0 \text{,}
\end{equation}
where $\psi(r)$ is a linear combination of functions $f^{lm}$, $h^{lm}$, $k^{lm}$ and $a^{lm}$. The potential $V(r)_{ele}$ appearing in
Eq. (\ref{wave2}) is
\begin{equation}\label{Ve}
V(r)_{ele}=f(r)\left[\frac{l(l+1)}{r^2}\right].
\end{equation}

We consider the effective potential equations of gravitational perturbation according to papers \cite{Kodama:2003jz} and \cite{Lopez-Ortega:2006aal}. 
Since it is difficult to calculate the scalar type of gravitational perturbation, only tensor and vector types of gravitational perturbation are calculated in this paper.
Their potential functions in four-dimensional spacetime are as follows:
\begin{equation}
V(r)_{grav (T)}= V(r)_{scalar} \text{,}
\end{equation}
\begin{equation}
V(r)_{grav (V)}=f(r)\left[\frac{l(l+1)}{r^2} -\frac{r}{2}f^{\prime \prime \prime}(r) \right] \text{.}
\end{equation}

\section{The QNMs in different fields}
\subsection{Three calculation methods}
The Mashhoon method is the P$\ddot{\text{o}}$schl-Teller potential approximation method \cite{Poschl:1933zz}, which uses the P$\ddot{\text{o}}$schl-Teller potential $V_{P T}$ to approximate the effective potential $V$ in the tortoise coordinate system.

The ``asymptotic iteration method'' (AIM) was applied to solve second order differential equations for the first time in \cite{Ciftci:2005xn}.
This method was then used to obtain the QNMs of field perturbation in Schwarzschild black hole \cite{Cho:2009cj}.

The Gundlach-Price-Pullin method, also known as the ``time-domain integration method'' or ``finite difference method'' \cite{Gundlach:1993tp}.
The time-domain profile can be obtained by this method.
Furthermore, we use the ``least square analysis'' (LSA) to extract the QNMs in the time-domain profile.
Specifically, the LSA is used to fit the attenuated linear regression equation, and then the approximate $\omega$ in time-domain profile are obtained.

There are specific explanations of this three methods in \cite{Yan:2020hga, Yan:2020nvk}, which will not be repeated here.

\subsection{Analysis of numerical results}
We use three different methods to calculate the QNMs of the charged non-commutative black hole and give the numerical results corresponding to different parameters in the scalar, electromagnetic and gravitational fields.
The WKB method is mostly used when calculating QNMs. 
However, due to the complexity of the metric of charged non-commutative black holes, it is very difficult to calculate with WKB.
In addition, the WKB method will make serious errors when calculating non-commutative black holes \cite{Yan:2020hga, Yan:2020nvk}.
Therefore, this paper chooses to use other methods for calculation, which can not only ensure the accuracy of the results, but also greatly shorten the calculation time.

\begin{table}[hbt]\centering\caption{The QNMs of charged non-commutative black holes in scalar, electromagnetic and gravitational fields are calculated by the Msahhoon method, AIM and time-domain integration method. The parameters $M=1$, $\theta=0.1$, $l=3$ and $n=0$ are selected.  $l$ is the multipole quantum number, $n$ is the overtone number.}
\label{NR1}
\begin{tabular*}{16cm}{*{4}{c @{\extracolsep\fill}}}	
\hline	
$Q$ & $\omega(\text{Mashhoon})$ & $\omega(\text{AIM})$ & $\omega(\text{Time-domain})$     \\
\cline{2-4}
&\multicolumn{3}{c}{$V(r)_{scalar, grav (T)}$}	 \\
\hline
0.0 & 0.678098 $-$ 0.0970906 $i$ & 0.675370 $-$ 0.0965014 $i$ & 0.675395 $-$ 0.0964935 $i$\\ 
0.1 & 0.679227 $-$ 0.0971423 $i$ & 0.676502 $-$ 0.0965541 $i$ & 0.676528 $-$ 0.0965400 $i$\\ 
0.2 & 0.682664 $-$ 0.0972957 $i$ & 0.679949 $-$ 0.0967101 $i$ & 0.679975 $-$ 0.0967086 $i$\\ 
0.3 & 0.688566 $-$ 0.0975444 $i$ & 0.685867 $-$ 0.0969634 $i$ & 0.685894 $-$ 0.0969463 $i$\\ 
0.4 & 0.697221 $-$ 0.0978746 $i$ & 0.694542 $-$ 0.0973012 $i$ & 0.694575 $-$ 0.0972955 $i$\\ 
\hline	
&\multicolumn{3}{c}{$V(r)_{ele}$} \\		
\cline{2-4}
0.0 & 0.659686 $-$ 0.0962250 $i$ & 0.656900 $-$ 0.0956190 $i$ & 0.656924 $-$ 0.0956027 $i$\\ 
0.1 & 0.660805 $-$ 0.0962783 $i$ & 0.658023 $-$ 0.0956732 $i$ & 0.658046 $-$ 0.0956660 $i$\\ 
0.2 & 0.664212 $-$ 0.0964366 $i$ & 0.661440 $-$ 0.0958341 $i$ & 0.661464 $-$ 0.0958337 $i$\\ 
0.3 & 0.670065 $-$ 0.0966937 $i$ & 0.667310 $-$ 0.0960958 $i$ & 0.667334 $-$ 0.0960802 $i$\\ 
0.4 & 0.678654 $-$ 0.0970363 $i$ & 0.675921 $-$ 0.0964457 $i$ & 0.675948 $-$ 0.0964483 $i$\\ 
\hline 
&\multicolumn{3}{c}{$V(r)_{grav (V)}$} \\		
\cline{2-4}
0.0 & 0.602419 $-$ 0.0933619 $i$ & 0.598781 $-$ 0.0927648 $i$ & 0.599109 $-$ 0.0923915 $i$\\ 
0.1 & 0.603694 $-$ 0.0934237 $i$ & 0.600072 $-$ 0.0929017 $i$ & 0.600379 $-$ 0.0924282 $i$\\ 
0.2 & 0.607583 $-$ 0.0936078 $i$ & 0.604036 $-$ 0.0933296 $i$ & 0.604256 $-$ 0.0926003 $i$\\ 
0.3 & 0.614294 $-$ 0.0939080 $i$ & 0.610970 $-$ 0.0940573 $i$ & 0.610945 $-$ 0.0928004 $i$\\ 
0.4 & 0.624215 $-$ 0.0943088 $i$ & 0.621410 $-$ 0.0948619 $i$ & 0.620833 $-$ 0.0930502 $i$\\ 
\hline 
\end{tabular*}
\end{table}

\begin{table}[hbt]\centering\caption{Similar to Table \ref{NR1}, but the parameter $Q=0.4$ is selected.}
\label{NR2}
\begin{tabular*}{16cm}{*{4}{c @{\extracolsep\fill}}}	
\hline	
$\theta$ & $\omega(\text{Mashhoon})$ & $\omega(\text{AIM})$ & $\omega(\text{Time-domain})$     \\
\cline{2-4}
&\multicolumn{3}{c}{$V(r)_{scalar, grav (T)}$}	 \\
\hline
0.05 & 0.697221 $-$ 0.0978747 $i$ & 0.694550 $-$ 0.0973022 $i$ & 0.694587 $-$ 0.0972957 $i$\\ 
0.10 & 0.697221 $-$ 0.0978746 $i$ & 0.694542 $-$ 0.0973012 $i$ & 0.694575 $-$ 0.0973032 $i$\\ 
0.15 & 0.697226 $-$ 0.0978462 $i$ & 0.694316 $-$ 0.0971677 $i$ & 0.694245 $-$ 0.0971885 $i$\\ 
0.20 & 0.697348 $-$ 0.0972941 $i$ & 0.693019 $-$ 0.0960055 $i$ & 0.693263 $-$ 0.0959335 $i$\\ 
\hline	
&\multicolumn{3}{c}{$V(r)_{ele}$} \\		
\cline{2-4}
0.05 & 0.678654 $-$ 0.0970364 $i$ & 0.675928 $-$ 0.0964458 $i$ & 0.675963 $-$ 0.0964321 $i$\\ 
0.10 & 0.678654 $-$ 0.0970363 $i$ & 0.675921 $-$ 0.0964457 $i$ & 0.675949 $-$ 0.0964380 $i$\\ 
0.15 & 0.678664 $-$ 0.0969893 $i$ & 0.675685 $-$ 0.0962744 $i$ & 0.675610 $-$ 0.0962781 $i$\\ 
0.20 & 0.678840 $-$ 0.0962754 $i$ & 0.674417 $-$ 0.0948607 $i$ & 0.674701 $-$ 0.0948216 $i$\\ 
\hline 
&\multicolumn{3}{c}{$V(r)_{grav (V)}$} \\		
\cline{2-4}
0.05 & 0.624208 $-$ 0.0943465 $i$ & 0.621324 $-$ 0.0937073 $i$ & 0.621351 $-$ 0.0936956 $i$\\ 
0.10 & 0.624215 $-$ 0.0943088 $i$ & 0.621410 $-$ 0.0948619 $i$ & 0.620834 $-$ 0.0930569 $i$\\ 
0.15 & 0.625440 $-$ 0.0888988 $i$ & 0.623528 $-$ 0.0977381 $i$ & 0.624553 $-$ 0.0846251 $i$\\ 
0.20 & 0.641806 $-$ 0.0712106 $i$ & 0.631588 $-$ 0.0838247 $i$ & 0.647061 $-$ 0.0775725 $i$\\ 
\hline 
\end{tabular*}
\end{table}

\begin{figure}[htbp]
\centering
\includegraphics[height=4cm,width=16cm]{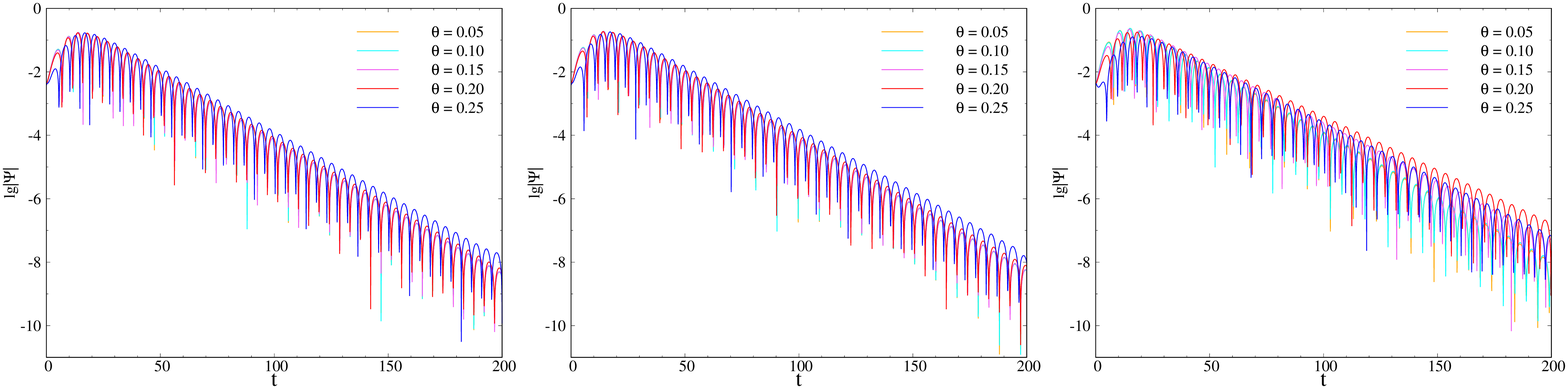}
\caption{Time-domain profiles of $V(r)_{scalar, grav (T)}$ (left), $V(r)_{ele}$ (middle) and $V(r)_{grav (V)}$ (right). The parameters $M=1$, $Q=0.4$, $l=3$, $n=0$, $v_{c}=10$ and $\sigma=3$ are selected. $v_{c}$ and $\sigma$ are the parameters of the input initial Gaussian wave.}\label{fig2}
\end{figure}

\begin{figure}[htbp]
\centering
\includegraphics[height=4cm,width=16cm]{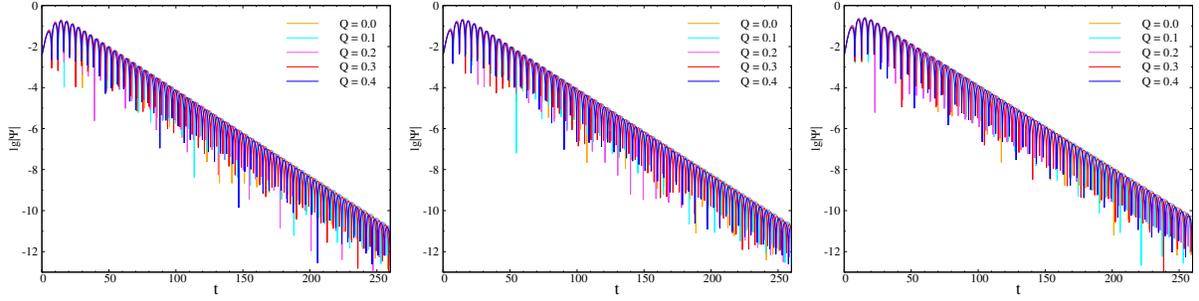}
\caption{Similar to Fig. \ref{fig2}, but the parameter $\theta=0.1$ is selected.}
\label{fig3}
\end{figure}

\begin{figure}[htbp]
\centering
\includegraphics[height=4cm,width=16cm]{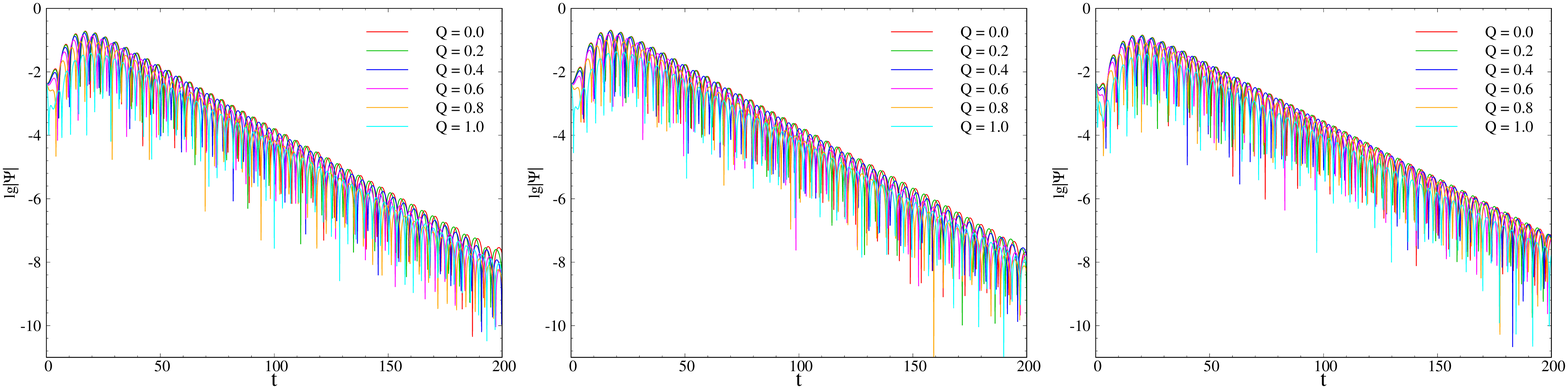}
\caption{Similar to Fig. \ref{fig2}, but the parameter $\theta=\theta_{max}$ is selected.}
\label{fig4}
\end{figure}

It can be seen from Table \ref{NR1} and Table \ref{NR2} that these three methods are effective, but the accuracy of QNMs calculated by the msahhoon method is low, while the numerical results obtained by AIM and the time-domain integration method are more accurate.
Moreover, the numerical results of QNMs change slightly with the increase of charge $Q$ and non-commutative parameter $\theta$ in scalar, electromagnetic and gravitational fields.
The change is so small that it can almost be ignored.
On the other hand, we also give the profiles of perturbation evolution with time in scalar, electromagnetic and gravitational fields corresponding to different parameters.
As shown in the Figs. \ref{fig2} and \ref{fig3}.
The case of a critical black hole is shown in Fig. \ref{fig4}.
It can be clearly seen that the damping of charged non-commutative black holes with time has the stability of dynamical evolution under any parameters.

\section{The shadow of a black hole}
\subsection{The photon orbits follow the null geodesics in black hole spacetime}
For the metric of Eq. (\ref{metricofsss}), the Lagrangian of null geodesics is described by
\begin{equation}\label{Lagrangian}
\mathcal{L}(x, \dot{x})=\frac{1}{2} g_{\mu \nu} \dot{x}^{\mu} \dot{x}^{\nu}.
\end{equation}
Substituting Eq. (\ref{metricofsss}) into Eq. (\ref{Lagrangian}), and for convenience, we only consider the geodesic on the equatorial plane, that is, $\vartheta= \pi/2$. Therefore, we obtain the following
\begin{equation}\label{LSSS}
\begin{aligned}
\mathcal{L}(x, \dot{x})  &= \frac{1}{2}\left( g_{00} \dot{t}^{2} +  g_{11} \dot{r}^{2} + g_{22} \dot{\vartheta}^{2} + g_{33} \dot{\varphi}^{2} \right) \\
   &= \frac{1}{2}\left[-A(r)  \dot{t}^{2} + B(r) \dot{r}^{2} + D(r)\dot{\vartheta}^{2} + D(r) \sin ^{2} \vartheta \dot{\varphi}^{2}\right]\\
   &= \frac{1}{2}\left[-A(r)  \dot{t}^{2}+B(r) \dot{r}^{2}+D(r)\dot{\varphi}^{2}\right] \\
\end{aligned}.
\end{equation}

The motion of photons is described by the Euler-Lagrange equation
\begin{equation}\label{E-L}
\frac{d}{d \lambda}\left(\frac{\partial \mathcal{L}}{\partial \dot{x}^{\mu}}\right)-\frac{\partial \mathcal{L}}{\partial x^{\mu}}=0,
\end{equation}
where $\lambda$ is the affine parameter, $\dot{x}^{\mu}$ represents the four-velocity components of light ray.
Substituting Eq. (\ref{LSSS}) into Eq. (\ref{E-L}),
\begin{equation}\label{con}
\begin{aligned}
&\frac{d}{d \lambda}\left(- A(r)  \dot{t} \right) =0 \\
&\frac{d}{d \lambda}\left( B(r) \dot{r} \right) =0 \\
&\frac{d}{d \lambda}\left( D(r) \dot{\varphi} \right) =0
\end{aligned}.
\end{equation}
Obviously, there exist two conserved quantities (constant of motion):
\begin{equation}\label{twoterm}
E=-\frac{\partial \mathcal{L}}{\partial \dot{t}}=A(r) \dot{t}, \quad L=\frac{\partial \mathcal{L}}{\partial \dot{\varphi}}=D(r) \dot{\varphi},
\end{equation}
which are defined as energy and angular momentum of photons, respectively.

We use a first integral of the geodesic equation, namely $\mathcal{L}(x, \dot{x})=0$ (for light).
Therefore, we obtain the equation as follows
\begin{equation}\label{LSSS90}
-A(r) \dot{t}^{2}+B(r) \dot{r}^{2}+D(r)\dot{\varphi}^{2} = 0,
\end{equation}
Using $\dot{r}^{2} / \dot{\varphi}^{2}=(d r / d \varphi)^{2}$ and Eq. (\ref{twoterm}), the orbit equation of the lightlike geodesic is given as follows:
\begin{equation}\label{varphi}
\left(\frac{d r}{d \varphi}\right)^{2}=\frac{D(r)}{B(r)}\left(\frac{D(r)}{A(r)}\cdot\frac{1}{b^{2}}-1\right),
\end{equation}
where $L/E=b$ is defined as the impact parameter, it is interpreted as the vertical distance between the geodesic line and the parallel line passing through the origin.

When a photon with a certain impact parameter $b$ is in the critical state of being captured or escaping by a black hole, the photon is in an unstable circular orbit near the black hole.
Many circular orbits form a photon sphere shell around the black hole.
This critical photon sphere radius is defined as $r_{c}$, and its corresponding critical impact parameter is defined as $b_{c}$.
The motion of light on the critical photon sphere must have a certain value, so
\begin{equation}\label{dotrddotr}
\dot{r}=0, \quad \ddot{r}=0,
\end{equation} 
should be satisfied. 
According to Eq. (\ref{LSSS90}), we give the expressions of $\dot{r}$ and $\ddot{r}$ as follows:
\begin{equation}
\dot{r} = \left( \frac{A(r) \dot{t}^{2} - D(r) \dot{\varphi}^{2}}{B(r)} \right)^{\frac{1}{2}},
\end{equation}
\begin{equation}
\begin{aligned}\label{ddotr}
\ddot{r} &= \frac{\partial r}{\partial  \lambda} \left[ \frac{\partial}{\partial r} \left( \dot{r} \right) \right] = \dot{r} \left( \frac{\partial \dot{r}}{\partial r} \right)  \\
&= \frac{\left(\dot{t}^{2} A^{\prime}(r) - \dot{\varphi}^{2} D^{\prime}(r)  \right)B(r) - \left(\dot{t}^{2} A(r) - \dot{\varphi}^{2}D(r) \right)B^{\prime}(r)}{2 B^{2}(r)} \\
\end{aligned},
\end{equation}
where `` $\prime$ '' is the derivative of $r$.
On the other hand, another condition of Eq. (\ref{con}) can be written as
\begin{equation}
\frac{d}{d \lambda}\left( B(r) \dot{r} \right) = \frac{dr}{d \lambda}\left[\frac{d}{d r}\left( B(r) \dot{r} \right)\right] = B^{\prime}(r) \dot{r}^{2} + B(r)\ddot{r} = 0,
\end{equation}
namely,
\begin{equation}\label{BB}
\left(\dot{t}^{2} A(r) - \dot{\varphi}^{2}D(r) \right)B^{\prime}(r) +\left(\dot{t}^{2} A^{\prime}(r) - \dot{\varphi}^{2} D^{\prime}(r)  \right)B(r) = 0.
\end{equation}
Furthermore, Eq. (\ref{ddotr}) can be simplified to
\begin{equation}
\ddot{r}=\frac{\dot{t}^{2} A^{\prime}(r) - \dot{\varphi}^{2} D^{\prime}(r)}{B(r)}.
\end{equation}
Therefore, Eq. (\ref{dotrddotr}) can be written as
\begin{equation}  
\left\{
  \begin{array}{l}
  A(r)  \dot{t}^{2} - D(r) \dot{\varphi}^{2} =0 \\
  \dot{t}^{2} A^{\prime}(r) - \dot{\varphi}^{2} D^{\prime}(r) =0  \\
  \end{array},
\right.
\end{equation}
and according to Eq. (\ref{twoterm}), the equations expressed by variable $b$ can be given as
\begin{equation}\label{EE}
\left\{
  \begin{array}{l}
  D(r) - A(r) b^{2} =0 \\
  D(r)^{2} A^{\prime}(r) - D^{\prime}(r) A(r)^{2} b^{2} =0  \\
  \end{array}.
\right.
\end{equation}
By solving the above equation, the critical photon sphere radius $r_{c}$ and critical impact parameter $b_{c}$ of any spherically symmetric black hole can be obtained.

It can be seen that the values of $r_{c}$ and  $b_{c}$ are independent of $B(r)$, but the photon orbits are affected by $B(r)$ at other positions (not in the critical photon orbit), as can be seen from Eq. (\ref{varphi}) and Eq. (\ref{EE}).

\subsection{The relationship between eikonal perturbation and shadow radius}
There is a deep relationship between the QNMs of a black hole and its shadow.
The relationship between the eikonal (short wavelengths or large multipole number) perturbation and the shadow radius of black holes in static spherically symmetric asymptotically flat spacetime is given in the paper \cite{Cardoso:2008bp}.
Recently, the relationship between the QNMs and the shadow of a rotating black hole has been given in \cite{Jusufi:2020dhz} and \cite{Yang:2021zqy}, respectively.

Since scalar, electromagnetic and gravitational perturbations of high-dimensional static black holes have the same behavior in the eikonal limit \cite{Kodama:2003jz, Ishibashi:2003ap, Kodama:2003kk}. Therefore, when the multipole number $l \rightarrow+\infty$, the general form of the potential function \cite{Cardoso:2008bp, Konoplya:2017wot, Churilova:2019jqx, Moura:2021eln} is expressed as
\begin{equation}
V_{eik}(r)=  \frac{f(r)}{r^{2}}l^{2} \text{.}
\end{equation}
In Einstein gravity, this limit can be applied to scalar, electromagnetic, Dirac and all types (tensor, vector, scalar) of gravitational perturbations.
However, when considering the coupled with nonlinear electromagnetic field \cite{Toshmatov:2019gxg, Toshmatov:2018tyo, Toshmatov:2018ell} and the modified gravitational theory perturbations \cite{Konoplya:2017wot}, there will be different situations.

The radius of the orbit is determined by
\begin{equation}
\left.\frac{\partial V_{eik}(r)}{\partial r}\right|_{r=r_{c}}=0
\text{,}
\end{equation}
therefore, it needs to satisfy
\begin{equation}\label{equrc}
2 f(r_{c})=r_{c} f^{\prime}(r_{c}) \text{.}
\end{equation}
The solution $r_{c}$ of the above equation is defined as the radius of the circular null geodesic, that is, the critical photon sphere orbit.

Under eikonal approximation, the QNMs of static spherically symmetric asymptotically flat spacetime in any dimension can be expressed as
\begin{equation}
\omega=l \Omega-i\left(n+\frac{1}{2}\right) \Lambda \text{.}
\end{equation}
The real part is determined by the angular velocity of the unstable null geodesic $\Omega$, and the imaginary part is determined by the Lyapunov exponent $\Lambda$. The expression \cite{Konoplya:2017wot, Moura:2021eln} is as follows:
\begin{equation}
\Omega=\frac{\sqrt{f\left(r_{c}\right)}}{r_{c}} \text{,}
\end{equation}
\begin{equation}
\Lambda=\sqrt{-\frac{r_{c}^{2}}{2 f\left(r_{c}\right)} \left[\frac{d^{2}}{d r_{\ast}^{2}} \left(\frac{f}{r^{2}}\right)\right]_{r=r_{c}}}=\sqrt{-\frac{r_{c}^{2}}{2}\left[f^{\prime}\left(\frac{f}{r^{2}}\right)^{\prime}+f\left(\frac{f}{r^{2}}\right)^{\prime\prime}\right]_{r=r_{c}}} \text{.}
\end{equation}
The shadow radius of a spherically symmetric black hole is given by
\begin{equation}\label{Rs}
R_{s}=\frac{1}{\Omega}=\frac{r_{c}}{\sqrt{f\left(r_{c}\right)}} \text{.}
\end{equation}
We can see that (\ref{Rs}) and (\ref{equrc}) combine to form Eq. (\ref{EE}).

In order to show this two-dimensional dark area, we use celestial coordinates $(\text{X},\text{Y})$ \cite{Shaikh:2019fpu, Lee:2021sws} to describe it, namely
\begin{equation}
R_{s} \equiv \sqrt{\text{X}^{2}+\text{Y}^{2}} \text{,}
\end{equation}
where
\begin{equation}
\text{X}=\lim _{r_{o} \rightarrow \infty}\left[-\left.r_{o}^{2} \sin \vartheta_{o} \frac{d \varphi}{d r}\right|_{\left(r_{o}, \vartheta_{o}\right)}\right] \text{,}
\quad 
\text{Y}=\lim _{r_{o} \rightarrow \infty}\left[\left.r_{o}^{2} \frac{d \vartheta}{d r}\right|_{\left(r_{o}, \vartheta_{o}\right)}\right] \text{,}
\end{equation}
and $(r_{o},\vartheta_{o})$ denotes the position coordinates of the observer at infinity, $\vartheta_{o}$ is the angle between the rotation axis of the black hole and the observer's line of sight.
The shadow of non-rotating spherically symmetric black holes is not affected by $\vartheta_{o}$.

\subsection{Discussion and evaluation of results}
\begin{figure}[htbp]
\centering
\includegraphics[height=7.5cm,width=15.5cm]{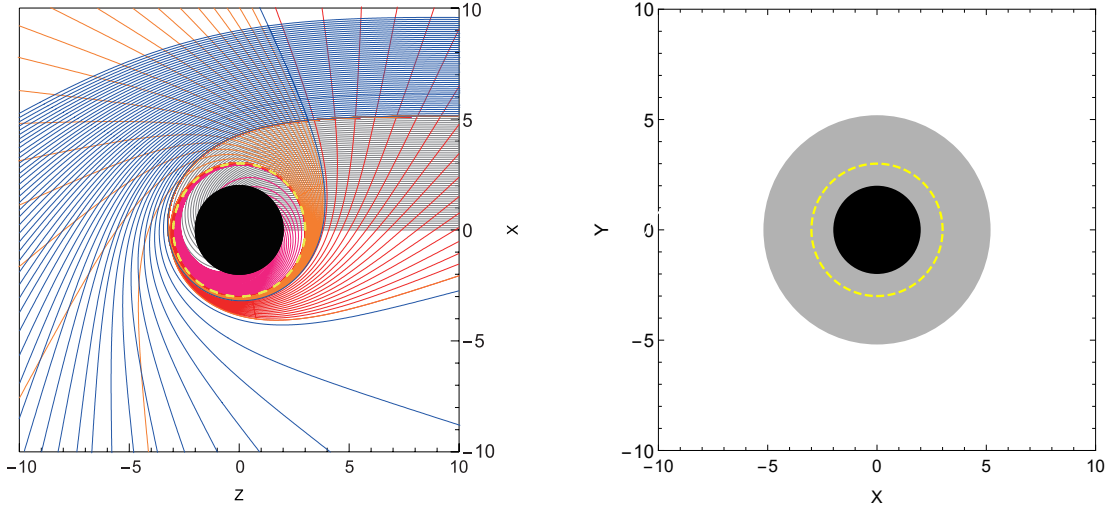}
\caption{The left panel shows the photon orbits around the Schwarzschild black hole on the equatorial plane ($\vartheta_{o}=0$).
Some photon trajectories do not circle the equatorial plane once (blue and gray), some trajectories circle the equatorial plane once (orange and magenta), and some trajectories circle the equatorial plane twice (red).
The right panel shows the event horizon $R_{EH}=2$ (the boundary of the black disc), the critical photon sphere orbit $r_{c}=3$ (the yellow dashed line) and the silhouette of shadow $b_{c}=R_{s}=3\sqrt{3}$ \cite{Perlick:2021aok, Gralla:2019xty} (the boundary of the gray disc) from the perspective of the vertical equatorial plane ($\vartheta_{o}= \pi /2$).
The parameter $M=1$ is selected.}
\label{fig5}
\end{figure}

First of all, Fig. \ref{fig5} is obtained by calculating Eq. (\ref{varphi}), which clearly shows the schematic diagram of black hole shadow, that is, the critical impact parameter $b_{c}$ is the shadow radius $R_{s}$.
By solving the solution of Eq. (\ref{EE}) with metric Eq. (\ref{Lapsefunction}), we can obtain $r_{c}$ and $b_{c}$ corresponding to different parameters $\theta$ and $Q$ in Table \ref{RCBC}.
It can be seen that $r_{c}$ and $b_{c}$ decrease with the increase of $\theta$ or $Q$.
\begin{table}[hbt]\centering\caption{Numerical results for $r_{c}$ and $b_{c}$ of charged non-commutative black holes. The parameter $M=1$ is selected.}
\label{RCBC}
\begin{tabular*}{16cm}{*{5}{c @{\extracolsep\fill}}}	
\hline	
&\multicolumn{2}{c}{$Q=0.0$} & \multicolumn{2}{c}{$Q=0.4$}\\
\cline{2-3} \cline{4-5}	
$\theta$ & $r_{c}$  & $b_{c}$  & $r_{c}$ & $b_{c}$  \\
\hline
0.10   & 2.999999956 & 5.196152418 & 2.889244213 & 5.052977266 \\ 
0.15   & 2.999955737 & 5.196145250 & 2.889131851 & 5.052957923 \\ 
0.20   & 2.998727753 & 5.195884333 & 2.886728313 & 5.052416616 \\ 
0.275811 & 2.980325376 & 5.190775292 & $\ast\ast\ast$ & $\ast\ast\ast$ \\
\hline	 
\end{tabular*}
\end{table}

\begin{figure}[htbp]
\centering
\includegraphics[height=11.2cm,width=15.5cm]{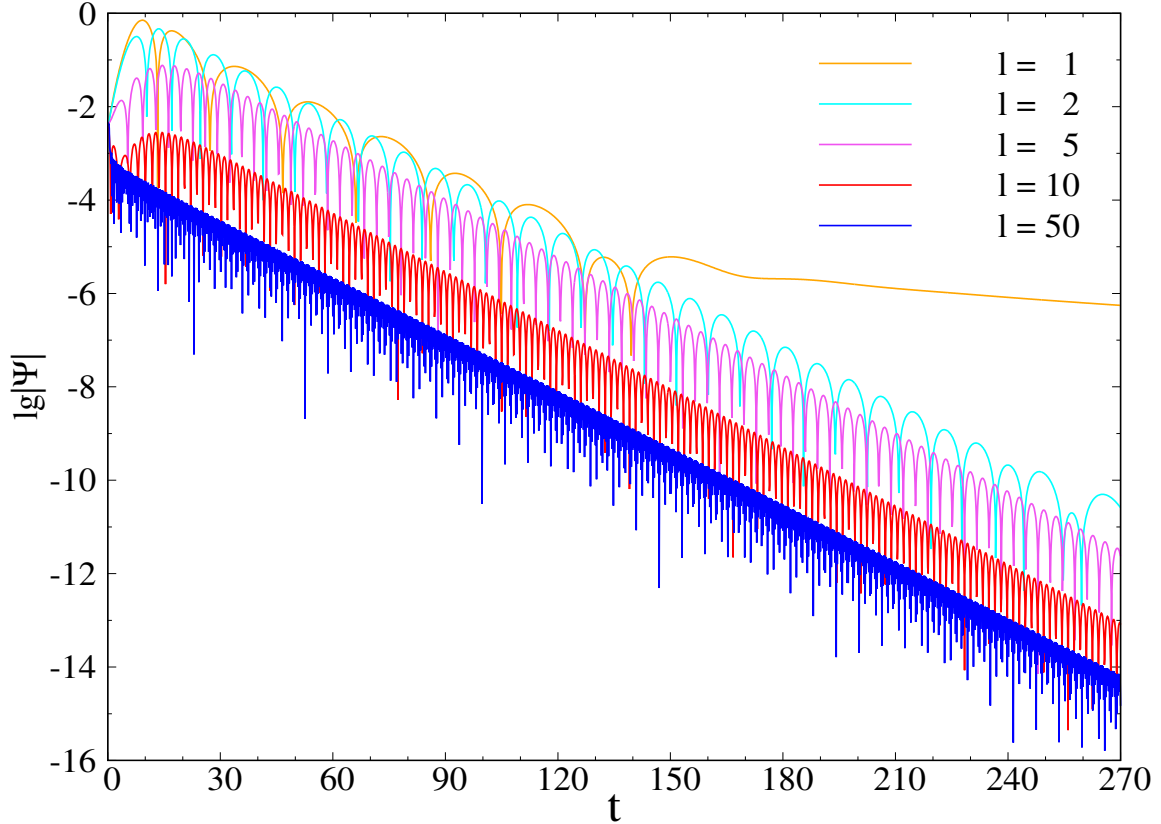}
\caption{Time-domain profile of perturbation approaching the eikonal limit ($l\gg1$). The parameters $M=1$, $\theta=0.1$, $Q=0.2$, $n=0$, $v_{c}=10$ and $\sigma=3$ are selected.}
\label{fig6}
\end{figure}

Next, we use the critical photon orbit method mentioned in the previous section to calculate the eikonal perturbation and compare the results with other methods.
Considering the time and accuracy of the calculation, we choose the AIM to calculate and compare with it. 
In addition, the ``relative deviation'' of the two methods is defined as \begin{equation}
\delta_{method}=\left|\frac{\omega(\text{AIM})-\omega(\Omega, \Lambda)}{\omega(\Omega, \Lambda)}\right| \text{.}
\end{equation}
The numerical results are shown in Table \ref{NR3}.
It can be seen that when the value of multipole number $l$ tends to be very large, the relative deviation $\delta_{method}$ is very small, which shows that the numerical results obtained by the two methods are very close, and it is also verified that the eikonal perturbation calculated by the critical photon orbit method is correct.
On the other hand, it is well known that when $l \rightarrow+\infty$, the imaginary part of QNMs tends to a fixed value, as also can be seen in Fig. \ref{fig6}.
\begin{table}[hbt]\centering\caption{The eikonal perturbation of charged non-commutative black hole is calculated by AIM and critical photon orbit method, and the relative deviations of the two numerical results are compared. The parameters $M=1$, $\theta=0.1$, $Q=0.2$ and $n=0$ are selected.}
\label{NR3}
\begin{tabular*}{16cm}{*{5}{c @{\extracolsep\fill}}}	
\hline	
&\multicolumn{2}{c}{$\text{QNMs}$} & \multicolumn{2}{c}{$\delta_{method}$}\\
\cline{2-3} \cline{4-5}	
$l$ & $\omega(\text{AIM})$  & $\omega(\Omega, \Lambda)$  & $\text{Re}(\omega)$ & $\text{Im}(\omega)$   \\
\hline
0   & 0.0803340 $-$ 0.254736 $i$ & 0.000000 $-$ 0.0964366 $i$ & $\infty$ & 1.64149 \\ 
1   & 0.158243 $-$ 0.0881689 $i$ & 0.193752 $-$ 0.0964366 $i$ & 0.224395 & 0.0937712\\ 
2   & 0.370671 $-$ 0.0947315 $i$ & 0.387504 $-$ 0.0964366 $i$ & 0.0454122 & 0.0179993\\ 
5   & 0.962009 $-$ 0.0961490 $i$ & 0.968760 $-$ 0.0964366 $i$ & 0.00701761 & 0.00299119\\ 
10  & 1.93414 $-$ 0.0963649 $i$ & 1.93752 $-$ 0.0964366 $i$ & 0.00174755 & 0.000744047\\   
50  & 9.68692 $-$ 0.0964337 $i$ & 9.68760 $-$ 0.0964366 $i$ & $7.01978 \times 10^{-5}$ & $3.00725 \times 10^{-5}$\\ 
$\vdots$ & $\vdots$ & $\vdots$ & $\vdots$ & $\vdots$ \\ 
100 & 19.3749 $-$ 0.0964367 $i$ & 19.3752 $-$ 0.0964366 $i$ & $1.54840 \times 10^{-5}$ & $1.03695 \times 10^{-6}$\\ 
\hline	 
\end{tabular*}
\end{table}

\begin{figure}[htbp]
\centering
\includegraphics[height=7.5cm,width=15.5cm]{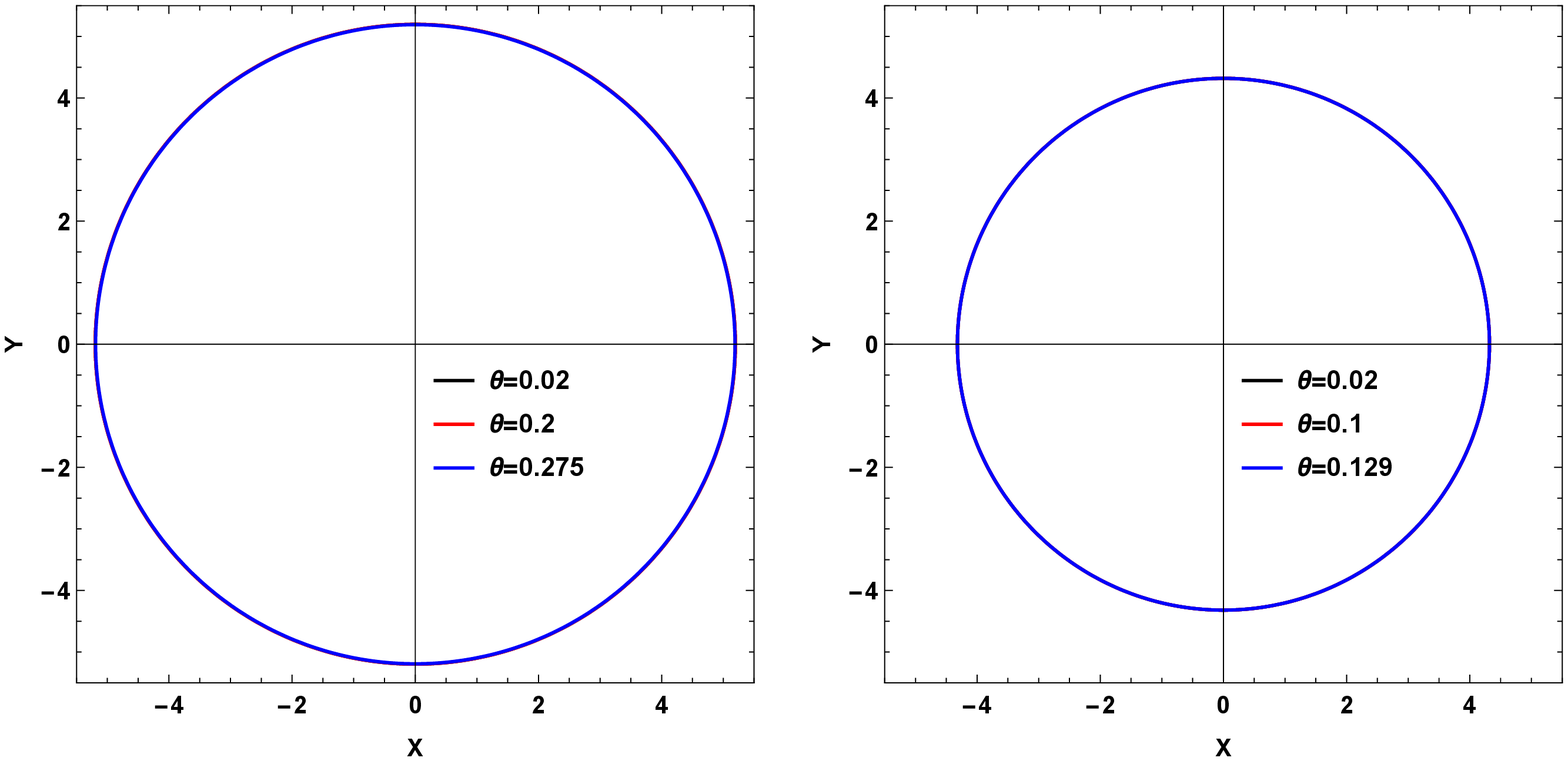}
\caption{The radius of shadow corresponding to different non-commutative parameter $\theta$. The parameters $Q=0$ (left panel), $Q=0.9$ (right panel), $M=1$ and $\vartheta_{o}= \pi /2$ are selected.}
\label{fig7}
\end{figure}

\begin{figure}[htbp]
\centering
\includegraphics[height=7.5cm,width=15.5cm]{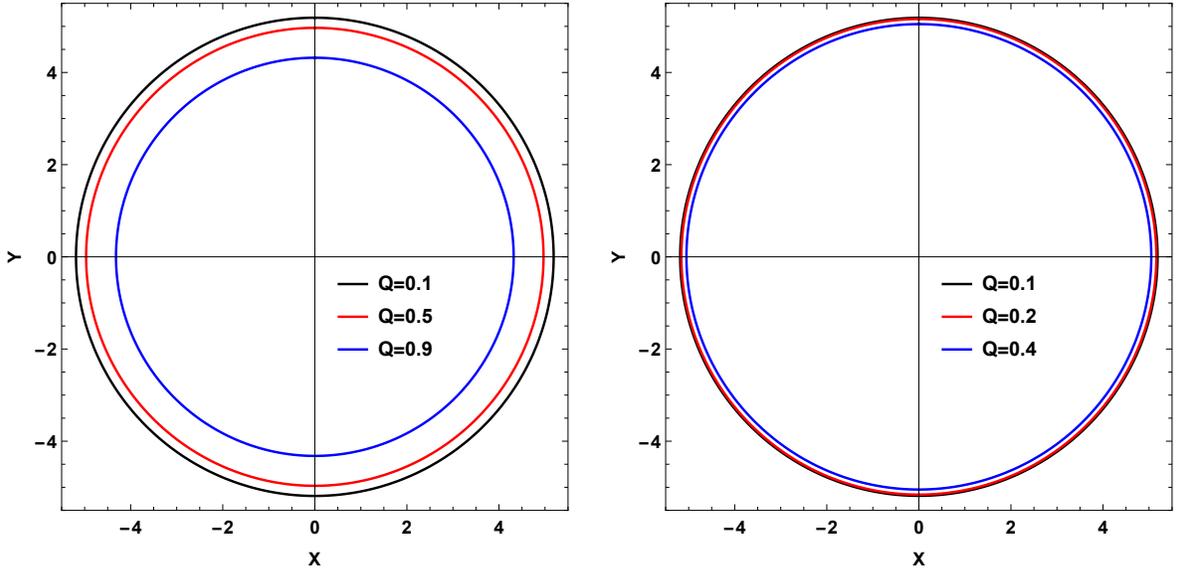}
\caption{The radius of shadow corresponding to different charge $Q$. The parameters $\theta=0.02$ (left panel), $\theta=0.25$ (right panel), $M=1$ and $\vartheta_{o}= \pi /2$ are selected.}
\label{fig8}
\end{figure}

And then, we obtain the shadow radius of some charged non-commutative black holes, as shown in Figs. \ref{fig7} and \ref{fig8}.
It can be seen from Fig. \ref{fig7} that when the black hole is uncharged and the charge is large, the influence of the non-commutativity on the shadow radius is not easy to distinguish.
Therefore, the $R_{s}$ decreases slightly when the $\theta$ increases.
In Fig. \ref{fig8}, in the case of small non-commutativity, the change of charge has an obvious effect on the shadow radius, which is due to the large change range of charge in the case of small non-commutativity. As shown in Fig. \ref{fig11}.

For spherically symmetric ($\tilde{r}_{\mathrm{sh}}$) and axisymmetric ($r_{\mathrm{sh}, A}$) black holes, the EHT gives the constraint range of the shadow radius of $\text{M87}^{\ast}$ \cite{EventHorizonTelescope:2021dqv, EventHorizonTelescope:2020qrl} and the constraint range of the shadow radius of $\text{Sgr\;A}^{\ast}$ from (Keck) and (VLTI) \cite{EventHorizonTelescope:2022xqj} as follows:
\begin{equation}\label{rangeofshadow}
4.31 M \approx r_{\text {sh,EHT-min}} \leq \tilde{r}_{\mathrm{sh}}, r_{\mathrm{sh}, A} \leq r_{\mathrm{sh}, \text { EHT-max}} \approx 6.08 M,
\end{equation}
\begin{equation}
\begin{aligned}\label{rangeofshadow2}
& 4.5 M \lesssim \tilde{r}_{\mathrm{sh}} \lesssim 5.5 M, \quad \text{(Keck)} \\
& 4.3 M \lesssim \tilde{r}_{\mathrm{sh}} \lesssim 5.3 M, \quad \text{(VLTI)} \\
\end{aligned},
\end{equation}
which has a confidence levels of $68 \%$ and $G=c=1$ has been set.
Therefore, we can constrain the parameters of the charged non-commutative black hole by using equations Eq. (\ref{rangeofshadow}) and Eq. (\ref{rangeofshadow2}).
\begin{figure}[htbp]
\centering
\includegraphics[height=12.5cm,width=15cm]{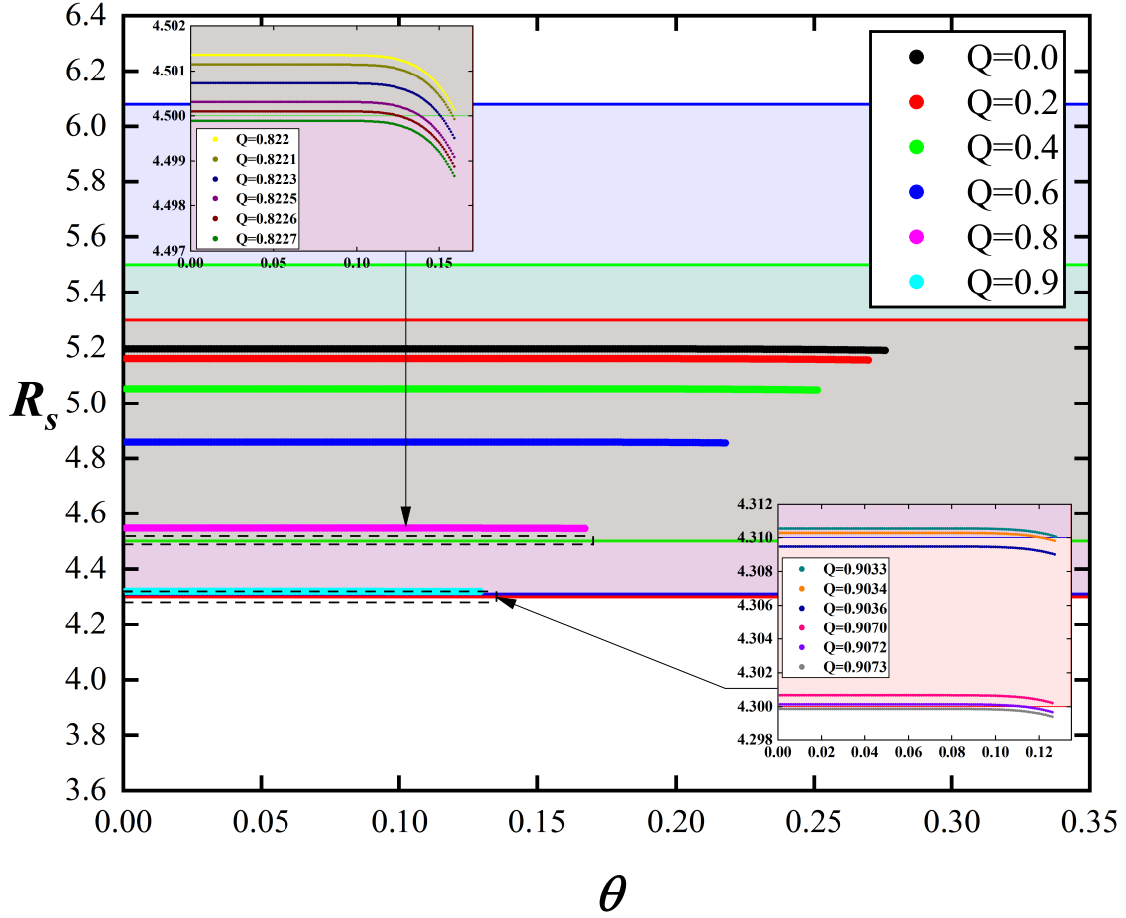}
\caption{The values of $R_{s}$ with the change of $\theta$ when $Q$ takes different values. The constraint range of $R_{s}$ by $\text{M87}^{\ast}$ (blue area), Keck (green area) and VLTI (red area). The parameter $M=1$ is selected.}
\label{fig9}
\end{figure}
In Fig. \ref{fig9}, we can see that when $Q$ is close to the critical maximum, the valid range of $\theta_{critical}$ will be less than $\theta_{max}$.
In other words, the valid range of $\theta$ will become smaller when $Q\rightarrow Q_{max}$,

We define the ``relative deviation'' between the shadow radius corresponding to different $\theta^{\prime}$ and the shadow radius in the commutative spacetime as
\begin{equation}
\delta_{R_{s}}=\left|\frac{R_{s}(\theta\rightarrow0)-R_{s}^{\prime}(\theta=\theta^{\prime})}{R_{s}^{\prime}(\theta=\theta^{\prime})}\right| \text{,}
\end{equation}
where $\theta^{\prime}$ satisfies the range $0<{\theta^{\prime}\leq\theta_{max}}$ corresponding to different $Q$ values.
Obviously, $\delta_{R_{s}}$ is a dimensionless quantity.
Therefore, it can be used to evaluate the deviation between non-commutative spacetime and commutative spacetime at the level of shadow radius.
The range of $\delta_{R_{s}}$ is shown in Fig. \ref{fig10}.
\begin{figure}[htbp]
\centering
\includegraphics[height=10cm,width=12.5cm]{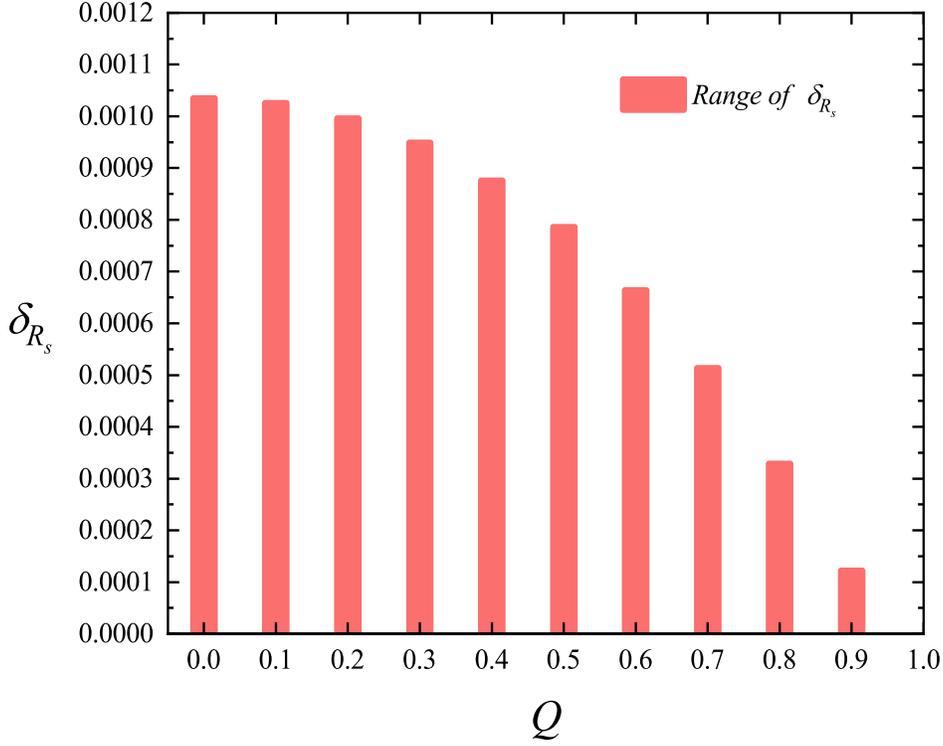}
\caption{The valid range of $\delta_{R_{s}}$ for various $Q$ values. The parameter $M=1$ is selected.}
\label{fig10}
\end{figure}

Therefore, we can find that the range of $\delta_{R_{s}}$ decreases with the increase of charge $Q$.
In other words, the non-commutativity of spacetime is more difficult to distinguish when the black hole carries more charge.

\section{Conclusions}
In this paper, different methods are used to calculate the perturbation of charged non-commutative black holes in scalar, electromagnetic and gravitational fields, and these numerical results are analyzed in detail. 
In addition, we have verified the relationship between the eikonal perturbation and the shadow radius in the non-vacuum Einstein's equation solution and calculated its shadow radius.
And then, we evaluated these results.
We give five conclusions as follows:

\begin{enumerate}

\item[1.]

In scalar, electromagnetic and gravitational fields, we obtain the accurate numerical results of QNMs in the charged non-commutative black hole spacetime.
The results obtained by AIM and time-domain integration method are more accurate than Msahhoon method, and the change of QNMs is almost be ignored with the increase of charge $Q$ and non-commutative parameter $\theta$.
In addition, the change of damping with time can ensure the stability of dynamical evolution under any parameter.

\item[2.]

We verified that the relationship between eikonal perturbation and shadow radius is valid in the case of non-vacuum Einstein's equation solution.

\item[3.]

We determined that the maximum value of $Q$, which comes from the constraint conditions (\ref{rangeofshadow}) and (\ref{rangeofshadow2}).
And when $Q$ approaches the maximum value, the effective range of $\theta$ will becomes smaller, that is, $\theta_{critical} < \theta_{max}$, as shown in Fig. \ref{fig9}.

\item[4.]

The shadow radius $R_{s}$ of a black hole decreases slightly as the non-commutative parameter $\theta$ increases, and the $R_{s}$ decreases as $Q$ increases.
When the black hole carries more charge, the range of $\delta_{R_{s}}$ value is smaller, which means that it is more difficult to distinguish the non-commutativity of spacetime by shadow radius.

\end{enumerate}

In addition, some papers have also discussed the shadow of non-commutative black holes. 
The conclusions of these papers are listed: 
1) 
The matter accretion rate increases rapidly with the increase of non-commutative parameter $\theta$ \cite{Gangopadhyay:2017tlp};
2)
In the low-frequency limit, the value of scattering/absorption cross section decreases with the increase of non-commutative parameter $\theta$ \cite{Anacleto:2019tdj};
3)
In paper \cite{Sharif:2016znp}, the shadow of a rotating charged non-commutative black hole is discussed, and it is found that the shadow is affected not only by non-commutative parameter $\theta$ and charge $Q$, but also by spin $a$ and the angle $\vartheta_{o}$. 
When the charge $Q$ increases, the deformation parameter $\delta_{s}$ of the silhouette of shadow decreases, that is, the shadow of the rotating black hole will maintain a circular shape with the increase of the charge $Q$. 
The particle orbit is also affected with the increase of $Q$.

\section*{Acknowledgements:}
The authors of this article thank developers of the AIM for their opening codes.

\section*{Data Availability Statement:} 
No data associated in the manuscript.

\bibliography{NC-RN}

\end{document}